# Analogy-Based Effort Estimation: A New Method to Discover Set of Analogies from Dataset Characteristics

Mohammad Azzeh
Department of Software Engineering
Applied Science University
Amman, Jordan POBOX 166
m.y.azzeh@asu.edu.jo

Ali Bou Nassif
Department of Computer Science
University of Western Ontario
London, Ontario, Canada, N6A 5B9
abounas@uwo.ca

**ABSTRACT**.

*Background*: Analogy-Based Effort Estimation (ABE) is one of the efficient methods for software effort estimation because of its outstanding performance and capability of handling noisy datasets.

*Problem & Objective:* Conventional ABE models usually use the same number of analogies for all projects in the datasets in order to make good estimates. Our claim is that using same number of analogies may produce overall best performance for the whole dataset but not necessarily best performance for each individual project. Therefore there is a need to better understand the dataset characteristics in order to discover the optimum set of analogies for each project rather than using a static *k* nearest projects.

*Method*: We propose a new technique based on Bisecting k-medoids clustering algorithm to come up with the best set of analogies for each individual project before making the prediction.

*Results & Conclusions*: With Bisecting k-medoids it is possible to better understand the dataset characteristic, and automatically find best set of analogies for each test project. Performance figures of the proposed estimation method are promising and better than those of other regular ABE models.

## 1. INTRODUCTION

Analogy Based Effort Estimation (ABE) is simplified a process of finding nearest analogies based on notion of retrieval by similarity [1, 12, 16, 24]. It was remarked that the predictive performance of ABE is a dataset dependent where each dataset requires different configurations and design decisions [14, 15, 19, 20]. Recent publications reported the importance of adjustment mechanism for generating better estimates in ABE than null-adjustment mechanism [1, 13, 26]. However, irrespective of the type of adjustment technique followed, the process of discovering the best set of analogies to be used is still a key challenge.

This paper focuses on the problem of how can we automatically come up with the optimum set of analogies for each individual project before making the prediction? Yet, there is no reliable method that can discover such set of nearest analogies before making prediction. Conventional ABE models start with one analogy and increase this number depending on the overall performance of the whole dataset then it uses the set of first *k* analogies that produces the best overall performance. However, a fixed *k* value that produces overall best performance does not necessarily provide the best performance for each individual project, and may not be suitable for other datasets. Our claim is that we can avoid sticking to a fixed best performing number of analogies which changes from dataset to dataset or even from a single project to another in the same dataset. Therefore we propose an alternative technique to tune ABE by proposing a Bisecting k-medoids (BK) clustering algorithm. The Bisecting procedure is used with k-medoids to avoid guessing number of clusters, by recursively applying the basic k-medoids algorithm and splitting each cluster into two sub-clusters to form a binary tree of clusters, starting from the whole dataset. This allows us to discover the structure of dataset efficiently and automatically come up with the best set of analogies as well as excluding irrelevant analogies for each individual test project. It is important to note that the discovered set of analogies does not necessarily include the same order of nearest analogies as in conventional ABE.

The rest of the article is structured as follows: Section 2 defines the research problem in more details. Section 3 provides the related work. Section 4 the methodology we propose to address the research problem. Section 5 presents the results we obtained. Section 6 presents discussion of our results and findings. Lastly Section 7 summarizes our conclusions and future work.

## 2. RESEARCH PROBLEM

Several studies in software effort estimation try to address the problem of finding optimum number of nearest analogies to be used by ABE [14, 15, 16, 31]. The conclusion drawn from these studies that using a static $k$ value that produces overall lowest *MMRE* does not necessarily provide the lowest *MRE* value for each individual project, and may not be suitable for other datasets. This shows that every dataset has different characteristics and this would have a significant impact on the process of discovering the best set of analogies. To illustrate our point of view and better understand this problem we carried out an intensive search to find the mean effort value of the nearest $k$ analogies that produces lowest *MRE* for every single test project as shown in Figure 1. For a dataset of size $n$ observations, the best $k$ value can range from 1 to $n-1$. Since a few number of datasets were enough to illustrate our viewpoint, we selected three datasets that vary in the size (i.e. one small dataset (Albrecht), one medium (Maxwell) and one large (Desharnais)). Figure 1 shows the bar chart of the best selected $k$ numbers for the three examined datasets, where x-axis represents project Id in that dataset and y-axis represents $k$ analogy number. It is clear that every single project favours different number of analogies. For example in Albrecht dataset, Three projects (id=3, 6, and 22) favoured $k$=15 which means that the final estimates for those project have been produced by using mean efforts of 15 nearest analogies. It is clear that there is no pattern for the process of $k$ selection. Therefore, using a fixed number of analogies for all test projects will far from optimum and there is provisional evidence that choosing the best set of analogies for each individual project is relatively subject to dataset structure.

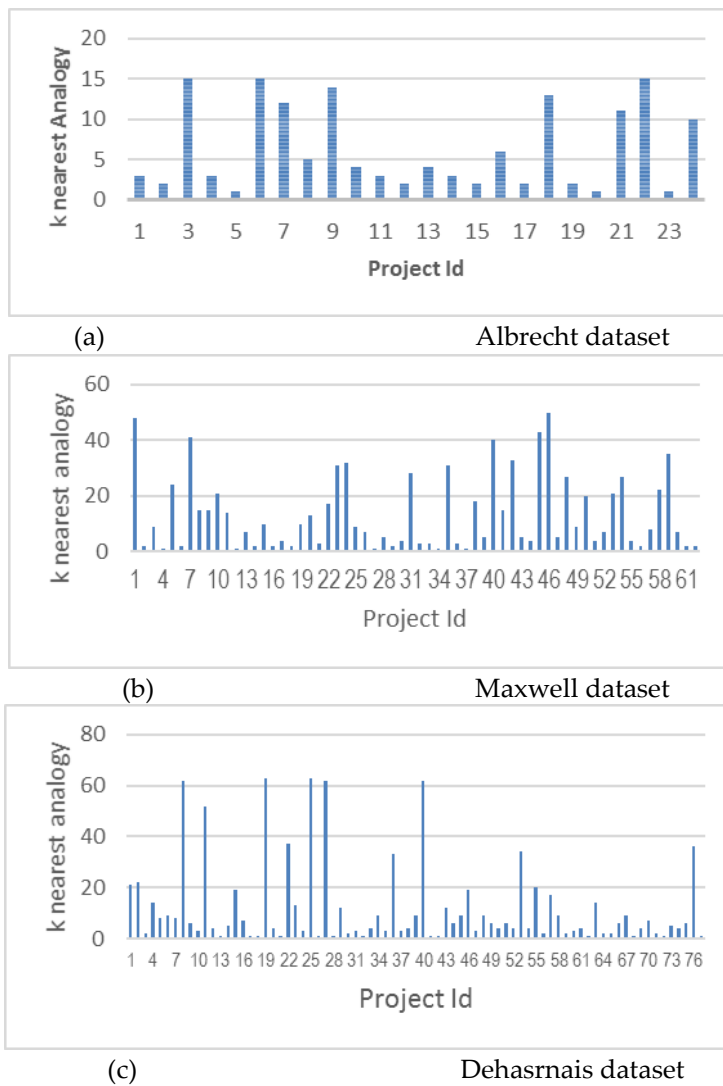

(a)                      Albrecht dataset

(b)                      Maxwell dataset

(c)                      Dehasrnais dataset

Figure 1. Bar chart of $k$ analogies for some datasets

## 3. RELATED WORKS

Software Effort Estimation is vital task for successful software project management [28, 29]. ABE method has been widely used for developing software effort estimation models based upon retrieval by similarity [2, 5, 10]. The data driven ABE method involves four primary steps [24]: (1) select $k$ nearest analogies using Euclidean distance function as depicted in Eq. 1. (2) Reuse efforts from the set of nearest analogies to find out effort of the new project. (3) Adjust the retrieved efforts to bring them closer to the new project. Finally, (4) retain the estimated project in the repository for future prediction.

$$d_{xy} = \frac{1}{m}\sqrt{\sum_{i=1}^{m}(x_i - y_i)^2} \qquad (1)$$

where $d_{xy}$ is the Euclidean distance between projects $x$ and $y$ across $m$ predictor features.

In spite of ABE generates better accuracy than other well-known prediction methods, it still requires adjusting the retrieved estimates to reflect the structure of nearest analogies on the final estimate [14]. Practically, the key factor of successful ABE method is finding the appropriate number of $k$ analogies. Several researchers [2, 5, 17, 13, 19] recommended using a fixed number of analogies starting from $k$=1 and increase this number until no further improvement on the accuracy can be obtained. This approach is somewhat simple, but not necessarily accurate, and relies heavily on the estimator intuitions [2]. In this direction, Kirsopp et al. [13] proposed making predictions from the $k$=2 nearest cases as it was found the best value for their investigated datasets. They have increased their accuracy values with case and feature subset selection strategies [13]. The conclusion can be drawn from their empirical studies is that the same $k$ number has been used for all datasets irrespective of their size and feature types (i.e. numerical, categorical and ordinal features). Azzeh [2] carried out an extensive replication study on various linear and non-linear adjustment strategies used in ABE in addition to finding the best $k$ number for these strategies. He found that $k$=1 was the most influential setting for all adjustment strategies over all datasets under investigation. On the other hand, Idri et al. [7] suggested using all projects that fall within a certain similarity threshold. They proposed a fuzzy similarity approach that can select the best analogies for which their similarity degrees are greater than the predefined threshold. This approach could ignore some useful projects which might contribute better when similarity between selected and unselected cases is negligible. Also the determination of the threshold value is a challenge on its own and needs expert intuition.

Another study focusing on $k$ analogies identification in the context of ABE is conducted by Li et al. [16]. They proposed a new model of ABE called AQUA which consists of two main phases: learning and prediction. During the learning phase, the model attempts to learn the $k$ analogies and best similarity threshold by performing cross-validation on all training projects. The obtained $k$ is then used during second phase to make prediction for different test projects. In their study Li et al. performed rigorous trials on actual and artificial datasets and they observed various effects of $k$ values.

Recently, Azzeh and Elsheikh [18] attempted to learn the $k$ value from the dataset characteristic. They applied the Bisecting k-medoid clustering algorithm on the historical datasets without using adjustment techniques or feature selection. The main observation was that while there is no optimum static $k$ value for all datasets, there is definitely a dynamic $k$ values for each dataset. However, the proposed approach has a significant limitation in which they used the un-weighted mean effort of the train projects of the leaf cluster whose medoid is closest to the test project to estimate the effort for that test project. Using such cluster does not ensure that all project in it are nearest analogies. In this paper we solve that problem by proposing a more robust approach in which in this study we focus mainly on discovering the optimum set of analogies rather than guessing only number of nearest analogies for each test project. Further, we want to investigate that whether the obtained set of analogies works well with different kinds of adjustment techniques. So we chose three well known adjustment techniques from the literature besides mean effort adjustment to investigate the potential improvements of using our model on the adjustment techniques. The techniques investigated in this study are:

1) *Similarity based adjustment*: This kind of adjustment aims to calibrate the retrieved effort values based on their similarity degrees with a target project. The general form of this technique involves sum of product of the normalized aggregated similarity degrees with retrieved effort value as shown in Eq. (2). Examples, on this approach, from literature are: AQUA [16], FGRA [2], and F-Analogy [7].

$$e_x = \frac{\sum_{i=1}^{k} SM(x, y_i) \times e_i}{\sum_{i=1}^{k} SM(x, y_i)} \quad (2)$$

Where $e_x$ and $e_i$ are the estimated effort and effort of $i^{th}$ source project respectively. *SM* is the normalized similarity degree between two projects (*SM*=1-*d*, where *d* is the normalized Euclidean distance obtained by Eq.1), and *k* is the number of analogies.

2) *Genetic Algorithm (GA) based Adjustment* [5]: this adjustment strategy uses GA to optimize the coefficient $\alpha_j$ for each feature distance based on minimizing *MMRE* as shown in Eq. (3). The main challenge with this technique is that it needs too many parameter configurations and user interactions such as chromosome encoding, mutation and crossover which makes replication is somewhat difficult task.

$$e_x = \frac{1}{k} \sum_{i=1}^{k} \left( e_i + \sum_{j=1}^{M} \alpha_j \times (f_{xj} - f_{ij}) \right) \quad (3)$$

where $f_{xj}$ is the $j^{th}$ feature value of the target project. $f_{ij}$ is the $j^{th}$ feature value of the nearest project $y_i$.

3) *Neural Network (NN) Based Adjustment* [17]: This technique attempts to learn the differences between effort values of target project and its analogies based on difference of their input feature values. These differences are then converted into the amount of change that will be added to the retrieved effort as shown in Eq. (4). The *NN* training function stops when *MSE* drops below the specified threshold= 0.01, and the model is trained based on Back-propagation algorithm.

$$e_x = \frac{1}{k} \sum_{i=1}^{k} (e_i + f(S_x, S_k)) \quad (4)$$

$f(S_x, S_k)$ is the neural network model. $S_x$ is the feature vector of a target project and $S_k$ is the feature matrix of the top analogies.

## 4 METHODOLOGY
### 4.1 The proposed Bisecting k-medoids algorithm:

The k-medoids [23] is a clustering algorithm related to the centroid-based algorithms which groups similar instances within a dataset into *N* clusters known a priori [11, 23, 27]. A medoid can be defined as the instance of a cluster, whose average dissimilarity to all the instances in the cluster is minimal i.e. it is a most centrally located point in the cluster. It is more robust to noise and outliers as compared to k-means because it minimizes the sum of pairwise dissimilarities instead of the sum of squared Euclidean distances [23]. The popularity of making use of k-medoids clustering is its ability to use arbitrary dissimilarity or distances functions, which also makes it an appealing choice of clustering method for software effort data as software effort datasets also exhibit very dissimilar characteristics. Since finding the suitable number of clusters is kind of guess [23] we employed bisecting procedure with k-medoids algorithm and propose Bisecting k-medoids algorithm (BK). BK is a variant of k-medoids algorithm that can produce hierarchical clustering by recursively applying the basic k-medoids. It starts by considering the whole dataset to be one cluster. At each step, one cluster is selected and bisected further into two sub clusters using the basic k-medoids as shown in the hypothetical example in Figure 2. Note that by recursively using a bisecting k-medoids clustering procedure, the dataset can be partitioned into any given number of clusters in which the so-obtained clusters are structured as a hierarchical binary tree. The decision whether to continue clustering or stop it depends on the comparison of variance degree between childes and their direct parent in the tree as shown in Eq. 5. If the maximum of variance of child clusters is smaller than variance of their direct parent then clustering is continued. Otherwise it is stopped and the parent cluster is considered as a leaf node. This criterion enables the BK to uniformly partition the dataset into homogenous clusters. To better understand the BK algorithm, we provide the pseudo code in Figure 3.

$$\text{Variance} = \frac{1}{n}\sum_{j=1, y_j \in C_i}^{n} \|y_j - v_i\|^2 \tag{5}$$

where $\|\cdot\|$ is the usual Euclidean norm, $y_j$ is the $j^{th}$ data object and $v_i$ is the centre of $i^{th}$ cluster ($C_i$). A smaller value of this measure indicates a high homogeneity (less scattering).

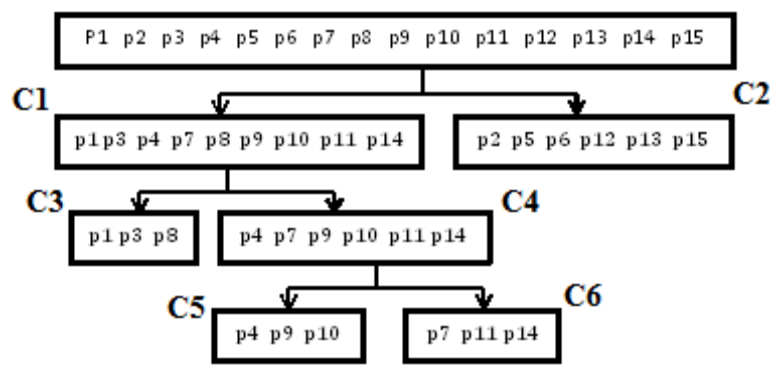

Figure 2 illustration of Bisecting k-medoids algorithm

```
    Input: The dataset X
 2  Output: The set of N clusters S={C₁, C₂, C₃, C₄, ..Cₙ}
 3  Initialization: Let V=X , S={}, NextLevl={}
 4  Repeat while size(V)> 0
 5    foreach Cluster C in V
 6      Comp ← variance (C)
 7      [C₁,C₂] ← k-medoids(C,2)
 8      Comp1 ← variance(C₁)
 9      Comp2 ← variance(C₂)
10    If(max(Comp1,Comp2)<Comp)
11      NextLevel ← NextLevel ∪ {C₁,C₂}
12    Else
13      S ← S ∪ {C}
14    End
15    V ← NextLevel
16    NextLevel ← {}
17  End
```

Figure 3 Bisecting k-medoids algorithm

## 4.2 The proposed *k*-ABE methodology

The proposed *k*-ABE model is described by the following steps:

1) For each new project, say *x*, we first cluster training datasets into *C* clusters using Bisecting k-medoids algorithm.

2) The Euclidian distance between project *x* and all training projects are computed. Then we sort all training projects according to their closeness to project *x*, smallest first.
3) Find first nearest neighbour from training dataset, Say *y*.
4) Find the cluster where project *y* belongs, say $C_y$.
5) Cluster the projects according to the distance values using Bisecting k-medoids algorithm as well. The cluster of most nearest projects ($C_n$) is selected.
6) The set of nearest projects is the intersection between clusters $C_y$ and $C_n$. this set will be the optimum set of analogies for the new project. In other words, i.e. we choose to use *k* many analogies for estimation. Compute Average of Effort values of all projects in that cluster using Eq. (6).

$$e_x = \frac{1}{k}\sum_{i=1}^{k} e_i \qquad (6)$$

## 4.3 Experimental Design

As it was reported in [14], most of the methods in literature were tested on a single or a very limited number of datasets, thereby reducing the credibility of the proposed method. To avoid this pitfall, we included 9 datasets from two different sources namely PROMISE [3] and ISBSG [8]. PROMISE is data repository consists of datasets donated by various researchers around the world. The datasets come from PROMISE are: albrecht, kemerer, cocomo, maxwell, desharnais, telecom, china and nasa. The albrecht dataset contains 24 software projects were developed by using third generation languages such as COBOL, PL1, etc. The dataset is described by 7 features: *input count, output count, query count, file count, line of code, function points* and *effort*. 18 projects were written in COBOL, 4 projects were written in PL1 and the rest were written in database management languages. The kemerer dataset consists of 15 projects described by 6 features for which two of them are categorical: *software, hardware* and 4 are continuous: *rawfp, ksloc, adjfp* and *effort*. Cocomo dataset consists of 63 software projects that are described by 17 features. The actual effort in the cocomo dataset is measured in person-months which represents the number of months that one person would need to develop a given project. The desharnais dataset consists of 81 software projects collected from Canadian software houses. This dataset is described by 11 features: *teamexp, managerexp, yearend, duration, transactions, entities, adjfp, adjfactor, rawfp, dev.env and effort*. The maxwell dataset is a relatively new dataset, which consists of 62 projects described by 23 features, collected from one of the biggest commercial banks in Finland. The dataset includes larger proportion of categorical features with 22 features which is hardly to be listed in this paper. Both telecom and nasa datasets are considered small size datasets with only 3 features each. China dataset is a very large dataset with 499 projects and 18 features, most of them are continuous. The other dataset comes from ISBSG data repository (release 10) [8] which is a large data repository consists of more than 4000 projects collected from different types of projects around the world. Since many projects have missing values only 500 projects with quality rating "A" are considered. 10 useful features were selected, 8 of which are numerical features and 2 of which are categorical features. The features used are: *AFP, input_count, output_count, enquiry_count, file_count, interface_count, add_count, delete_count, changed_count and effort.*

The descriptive statistics of such datasets are summarized in Table 1. From the table, we can conclude that datasets in the area of software effort estimation share relatively common characteristics [17]. They often have a limited number of observations that are affected by multicollinearity and outliers. We can also observe that all the datasets have positive skewness values which range from 1.78 to 4.36. This observation indicates that the datasets are extremely heterogeneous, which make sure that we test the proposed model adequately.

For each dataset we follow the same testing strategy, we used Leave-one-out cross validation where in each run, one project is selected as test and the remaining projects as training set. This procedure is performed until all projects within dataset are used as test projects. In each run, The prediction accuracy of different techniques is assessed using *MMRE, pred(0.25)* performance measure [4, 25]. *MMRE* computes mean of the absolute percentage of error between actual and predicted project effort values as shown in Eq. 7, 8. *pred(0.25)* is used as a complementary criterion to count the percentage of MREs that fall within less than 0.25 of the actual values as shown in Eq. 9.

Table 1. Statistical properties of the employed datasets

| Dataset | Feature | Size | Effort Data | | | | | |
|---|---|---|---|---|---|---|---|---|
| | | | unit | min | max | mean | median | skew |
| albrecht | 7 | 24 | months | 1 | 105 | 22 | 12 | 2.2 |
| kemerer | 7 | 15 | months | 23.2 | 1107.3 | 219.2 | 130.3 | 2.76 |
| nasa | 3 | 18 | months | 5 | 138.3 | 49.47 | 26.5 | 0.57 |
| ISBSG | 10 | 505 | hours | 668 | 14938 | 2828.45 | 1634 | 2.1 |
| desharnais | 11 | 77 | hours | 546 | 23940 | 5046 | 3647 | 2.0 |
| cocomo | 17 | 63 | months | 6 | 11400 | 683 | 98 | 4.4 |
| china | 18 | 499 | hours | 26 | 54620 | 3921 | 1829 | 3.92 |
| maxwell | 27 | 62 | hours | 583 | 63694 | 8223.2 | 5189.5 | 3.26 |
| telecom | 3 | 18 | months | 23.54 | 1115.5 | 284.33 | 222.53 | 1.78 |

$$MRE_i = \frac{|e_i - \hat{e}_i|}{e_i} \qquad (7)$$

$$MMRE = \frac{1}{N}\sum_{i=1}^{N} MRE_i \qquad (8)$$

where $e_i$ and $\hat{e}_i$ are the actual value and predicted values of $i^{th}$ project, and $N$ is the number of observations.

$$pred(0.25) = \frac{100}{N} \times \sum_{i=1}^{N} \begin{cases} 1 & if\ MRE_i \leq 0.25 \\ 0 & otherwise \end{cases} \qquad (9)$$

The Boxplot of absolute residuals and Wilcoxon sum rank test are also used to compare between different methods. The reason behind using these tests is because all absolute residuals for all models used in this study were not normally distributed. In turn, the obtained results from the proposed approach have benchmarked to other regular ABE models that use a fixed number of *k* analogies. In addition to that we used *win-tie-loss* algorithm [14] to compare the performance of *k*-ABE to other regular ABE models. To do so, we first check if two methods $M_i$; $M_j$ are statistically different according to the Wilcoxon test; otherwise we increase $tie_i$ and $tie_j$. If the distributions are statistically different, we update $win_i$; $win_j$ and $loss_i$; $loss_j$, after checking which one is better according to the performance measure at hand *E*. The performance measures used here are *MRE*, *MMRE*, median of *MRE* (*MdMRE*) and *pred*.

```
1    win_i=0,tie_i=0,loss_i=0
2    win_j=0,tie_j=0;loss_j=0
3    if WILCOXON(MRE(M_i), MRE(M_j), 95) says they are the same then
4            tie_i = tie_i + 1;
5            tie_j = tie_j + 1;
6    else
7            if better(E(M_i), E(M_j)) then
8                    win_i = win_i + 1
9                    loss_j = loss_j + 1
10           else
11                   win_j = win_j + 1
12                   loss_i = loss_i + 1
13           end if
14   end if
```

Figure 4. Pseudo code for *win-tie-loss* calculation between method $M_i$ and $M_j$ based on performance measure *E* [14].

# 5 RESULTS

In this paper we proposed Bisecting k-medoids algorithm to automatically come up with the optimum set of *k* analogies for each project based on analysing the characteristics of a dataset. To demonstrate that, we executed *k*-ABE over all investigated datasets and recorded the best obtained set for every test project. Figure 5 shows the relationship between numbers of *k* analogies sorted from 1 to 100 (please note that we took a part of results) and numbers of the projects selected these *k* values over all datasets. The x-axis represents *k* nearest analogies for the first 100 analogies, and y-axis represents number of projects selected a particular *k* value. The variability of *k* values demonstrates the capability of *k*-ABE model to dynamically discovering the different *k* analogies for individual projects that take into account the characteristics of each dataset. Furthermore, the procedure of selecting has become easier than first (i.e. where the estimator intuition was heavily used to choose the optimum number of analogy) since the entire best *k* selection process has been left to the BK.

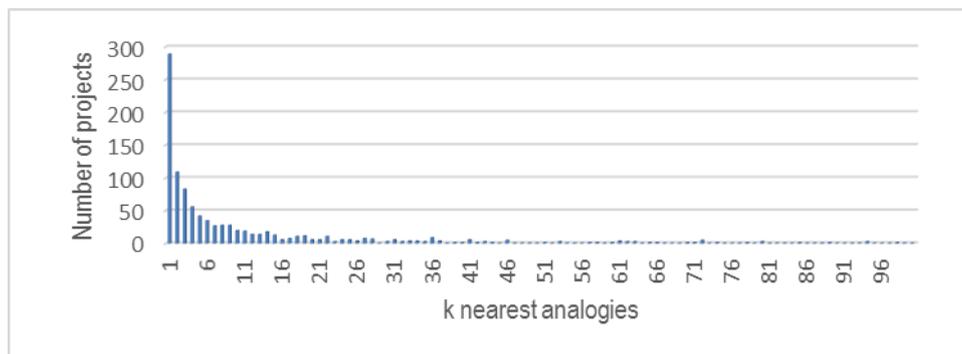

Figure 5. the relationship between number of projects and their associated *k* nearest analogies over all investigated datasets

For the sake of comparison we used the common ABE models that use fixed *k* value for all test instances. For example ABE1 represents the ABE variant that uses only the first nearest analogy, ABE2 represents the ABE variant that uses mean of the nearest two analogies and so forth. Apart from being able to identify optimum set of analogies for each test instance, the *k*-ABE method outperforms all the other regular ABE models as can be seen in Table 2. When we look at the *MMRE* values, we can see that in all nine datasets, *k*-ABE has never been outperformed by other methods with lowest *MMRE* values. This suggests that *k*-ABE has attained better predictive performance values than all other regular ABE models. This also shows the capability of BK to support small-size datasets such as in Kemerer and Albrecht. However, although it proved inaccurate in this study, the strategy of using fixed *k*-analogy may be appropriate in situations where a potential analogues and target project are similar in size feature and other effort drivers. On the other hand, There may be little basis for believing that either increasing or decreasing the *k*-analogies effort values of ABE models does not improve the accuracy of the estimation. However, overall results from Tables 2 and 3 revealed that there is reasonable believe that using dynamic *k*-analogies for every test project has potential to improve prediction accuracy of ABE in terms of *pred*. Concerning discontinuities in the dataset structure, there is clear evidence that the BK technique has capability to group similar projects together in the same cluster as appeared in the results of Maxwell, COCOMO, Kemerer and ISBSG.

Table 2 *MMRE* results of ABE variants

| Dataset | *k*-ABE | ABE1 | ABE2 | ABE3 | ABE4 | ABE5 |
|---|---|---|---|---|---|---|
| Albrecht | **30.5** | 71.0 | 66.5 | 77.8 | 73.9 | 72.4 |
| Kemerer | **38.2** | 55.9 | 77.7 | 77.4 | 86.2 | 86.0 |
| Desharnais | **34.3** | 60.1 | 51.5 | 50.0 | 50.2 | 50.0 |
| COCOMO | **29.3** | 157.1 | 363.2 | 350.4 | 327.3 | 325.2 |
| Maxwell | **27.7** | 182.6 | 132.7 | 120.6 | 149.3 | 144.0 |
| China | **31.6** | 45.2 | 44.2 | 46.7 | 48.5 | 51.7 |
| Telecom | **30.4** | 60.0 | 45.3 | 62.5 | 77.4 | 89.5 |
| ISBSG | **34.2** | 72.6 | 73.1 | 74.0 | 74.2 | 72.8 |
| NASA | **25.6** | 81.2 | 97.5 | 88.5 | 77.6 | 71.1 |

Table 3 *pred*(0.25) results of ABE variants

| Dataset | k-ABE | ABE1 | ABE2 | ABE3 | ABE4 | ABE5 |
|---|---|---|---|---|---|---|
| Albrecht | **41.7** | 29.2 | 33.3 | 33.3 | 37.5 | 41.7 |
| Kemerer | 26.7 | **40.0** | 20.0 | 20.0 | 13.3 | 20.0 |
| Desharnais | **40.3** | 31.2 | 31.2 | 37.7 | 37.7 | 39.0 |
| COCOMO | **57.1** | 12.7 | 19.0 | 19.0 | 15.9 | 22.2 |
| Maxwell | **56.5** | 9.7 | 19.4 | 17.7 | 14.5 | 17.7 |
| China | **46.7** | 38.3 | 43.5 | 43.3 | 41.9 | 39.7 |
| Telecom | **55.6** | 33.3 | 50.0 | 38.9 | 44.4 | 22.2 |
| ISBSG | 38.8 | **39.6** | 30.7 | 30.9 | 29.7 | 26.5 |
| NASA | **44.4** | 33.3 | 38.9 | 44.4 | 22.2 | 22.2 |

The variants of ABE methods are also compared using Wilcoxon sum rank test. The results of Wilcoxon sum rank test of absolute residuals are presented in Table 4. The solid black square indicates that there is significance difference between *k*-ABE and the variant under investigation. Predictions based on *k*-ABE model presented statistically significant and accurate estimations than others, confirmed by the results of *MMRE* as shown in Table 2. Except for small datasets such as Albrecht, Kemerer, Telecom and NASA, the statistical test results demonstrate that there are significant differences if the predictions generated by any *k*-ABE and other regular ABE models. So it seems that the small datasets are the most challenging ones because they have relatively small number of instances and large degree of heterogeneity between projects. This makes difficult to obtain a cluster of sufficient number of instances.

Table 4. Wilcoxon sum rank test results between *k*-ABE and other ABE variants

| Dataset | ABE1 | ABE2 | ABE3 | ABE4 | ABE5 |
|---|---|---|---|---|---|
| Albrecht |  |  |  |  |  |
| Kemerer |  |  |  |  |  |
| Desharnais | ■ | ■ | ■ | ■ | ■ |
| COCOMO | ■ | ■ | ■ | ■ | ■ |
| Maxwell | ■ | ■ | ■ | ■ | ■ |
| China | ■ | ■ | ■ | ■ | ■ |
| Telecom |  |  |  |  |  |
| ISBSG |  | ■ | ■ | ■ | ■ |
| NASA |  |  | ■ |  | ■ |

The *win-tie-loss* results in Table 5 shows that *k*-ABE outperformed regular ABE models with *win-loss*=102. Also these results are confirmed by the Boxplots of absolute residuals in Figure 6 which demonstrates that *k*-ABE has lowest median values and small box length than other methods for most datasets.

Table 5. Win-Tie-Loss Results of ABE variants

| ABE variant | win | tie | loss | win-loss |
|---|---|---|---|---|
| **k-ABE** | **108** | **21** | **6** | **102** |
| ABE1 | 64 | 45 | 26 | 38 |
| ABE2 | 21 | 52 | 62 | -41 |
| ABE3 | 29 | 59 | 47 | -18 |
| ABE4 | 16 | 39 | 80 | -64 |
| ABE5 | 13 | 38 | 84 | -71 |

The obtained performance figures raised a question concerning the efficiency of applying adjustment techniques to *k*-ABE. To answer this question we carried out an empirical study on the employed datasets, using three well known adjustment techniques: Similarity based adjustment, *GA* based adjustment and *NN* based adjustment in addition to the proposed BK technique. Their corresponding *k*-ABE variants are denoted by *k*-ABE$_{SM}$, *k*-ABE$_{GA}$ and *k*-ABE$_{NN}$ respectively. The obtained performance figures in terms of *MMRE* and *pred(0.25)* are recorded in Tables 8 and 9. In general there is no significant difference when applying various adjustment techniques than basic *k*-ABE. One possible reason is due to small number of instances in some clusters. It is well known that both *GA* and *NN* models need sufficient number of instances in order to produce good results; however this may not suitable for small datasets such as Albrecht, Kemerer, NASA, and Telecom. In contrast, there are little improvements on the accuracy when applying adjustment techniques than other regular ABE models for some datasets especially large ones.

Table 6. MMRE results of *k*-ABE variants

| Dataset | *k*-ABE | *k*-ABE$_{SM}$ | *k*-ABE$_{GA}$ | *k*-ABE$_{NN}$ |
|---|---|---|---|---|
| Albrecht | **30.5** | 59.7 | 94.8 | 71.7 |
| Kemerer | **38.2** | 45.7 | 52.8 | 72.6 |
| Desharnais | **34.3** | 40.2 | 47.2 | 100.5 |
| COCOMO | **29.3** | 73.6 | 70.1 | 118.5 |
| Maxwell | **27.7** | 55.2 | 54.7 | 60.3 |
| China | **31.6** | 61.4 | 64.6 | 76.0 |
| Telecom | **30.4** | 58.4 | 59.8 | 80.4 |
| ISBSG | **34.2** | 47.7 | 48.0 | 85.8 |
| NASA | **25.6** | 76.7 | 31.8 | 51.9 |

Table 7. *pred(0.25)* results of *k*-ABE variants

| Dataset | *k*-ABE | *k*-ABE$_{SM}$ | *k*-ABE$_{GA}$ | *k*-ABE$_{NN}$ |
|---|---|---|---|---|
| Albrecht | **41.7** | 16.7 | 16.7 | 8.3 |
| Kemerer | 26.7 | **33.3** | 20.0 | 13.3 |
| Desharnais | **40.3** | 29.9 | 20.8 | 14.3 |
| COCOMO | **57.1** | 11.1 | 9.5 | 7.9 |
| Maxwell | **56.5** | 22.6 | 21.0 | 22.6 |
| China | **46.7** | 13.0 | 11.8 | 12.6 |
| Telecom | **55.6** | 27.8 | 22.2 | 5.6 |
| ISBSG | **38.8** | 22.2 | 23.2 | 15.2 |
| NASA | **44.4** | 5.6 | 38.9 | 22.2 |

On the other hand, when comparing adjustment techniques to the basic *k*-ABE model we can notice that there is substantial improvement on the accuracy for all datasets except small ones. The statistical significant test in Table 8 shows that in general there is significance difference between the results of *k*-ABE and all other adjustment techniques: *k*-ABE$_{SM}$, *k*-ABE$_{GA}$ and *k*-ABE$_{NN}$. This suggests that the predictions generated by *k*-ABE are different than that of other adjustment techniques.

Table 8. Wilcoxon sum rank test results between *k*-ABE variants

| Dataset | *k*-ABE Vs. *k*-ABE$_{SM}$ | *k*-ABE Vs. *k*-ABE$_{GA}$ | *k*-ABE Vs. *k*-ABE$_{NN}$ | *k*-ABE$_{SM}$ Vs. *k*-ABE$_{GA}$ | *k*-ABE$_{SM}$ Vs. *k*-ABE$_{NN}$ | *k*-ABE$_{GA}$ Vs. *k*-ABE$_{NN}$ |
|---|---|---|---|---|---|---|
| Albrecht | ■ | ■ | ■ | | | |
| Kemerer | | | ■ | | | |
| Desharnais | ■ | ■ | ■ | | ■ | ■ |
| COCOMO | ■ | ■ | ■ | | | |
| Maxwell | ■ | ■ | ■ | | | |
| China | ■ | ■ | ■ | | | |
| Telecom | | | ■ | | | ■ |
| ISBSG | ■ | ■ | ■ | | ■ | ■ |
| NASA | ■ | | ■ | ■ | | |

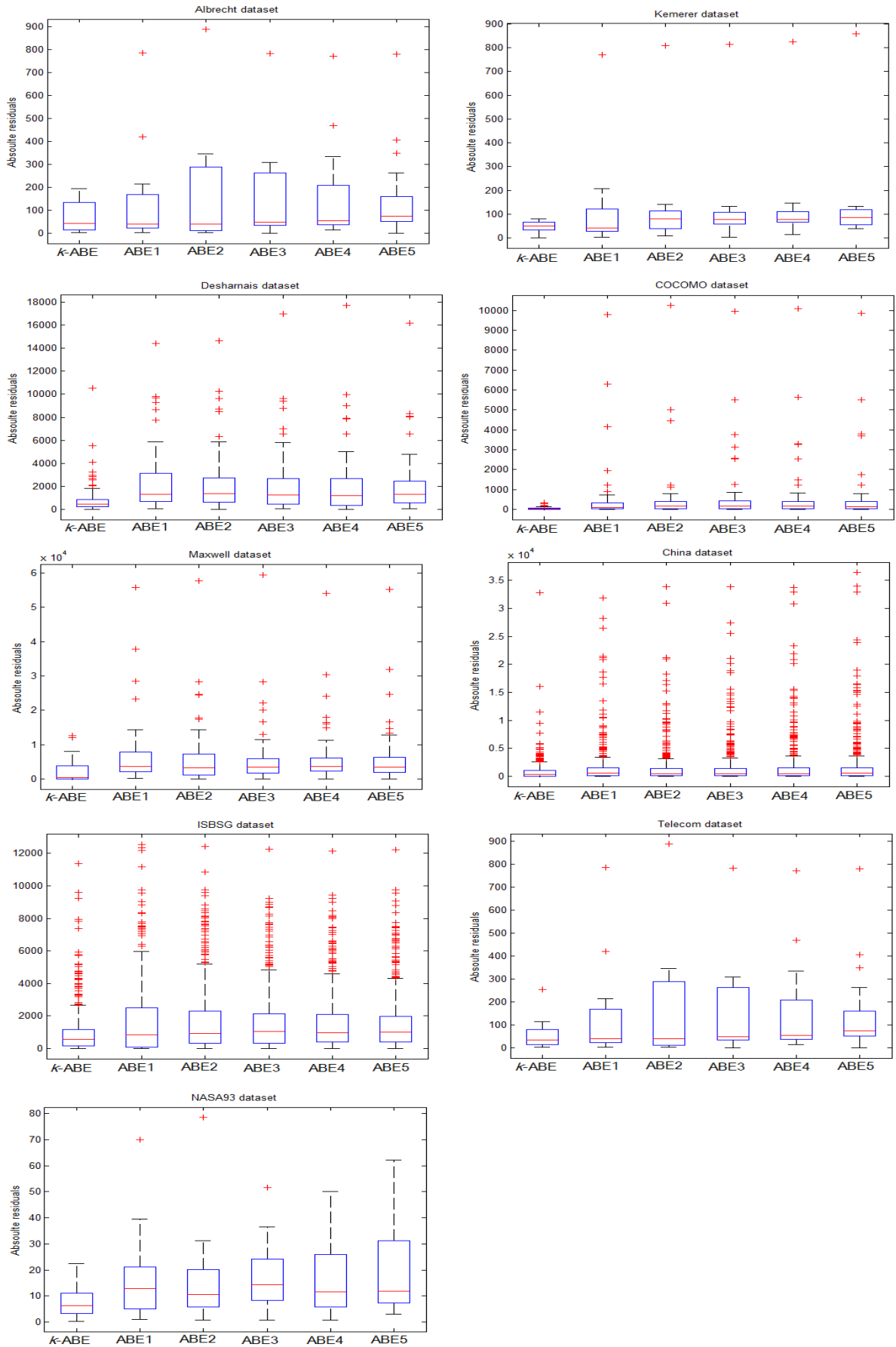

Figure 6. Boxplots of absolute residuals for ABE variants

Table 9. *win-tie-loss* results of *k*-ABE variants

| ABE variant | win | tie | loss | win-loss |
|---|---|---|---|---|
| *k*-ABE | 77 | 4 | 0 | 77 |
| *k*-ABE$_{SM}$ | 19 | 25 | 37 | -18 |
| *k*-ABE$_{GA}$ | 19 | 31 | 31 | -12 |
| *k*-ABE$_{NN}$ | 13 | 20 | 48 | -35 |

Boxplots of absolute residuals in Figure 7 show that there is significant difference between *k*-ABE and all other variants of *k*-ABE. The Boxplots suggest that:

1. All median values of *k*-ABE are very close to zero, indicating that the estimates were biased towards the minimum value where they have tighter spread. The median and range of absolute residuals of *k*-ABE are small, which revealed that at least half of the predictions of *k*-ABE are accurate than other variants. The box of *k*-ABE overlays the lower tail especially for Albrecht, COCOMO, Maxwell and China datasets, which also presents accurate prediction.

2. Although the number of outliers for ISBSG and China datasets is fairly high comparing to other datasets, they are not extremes like other variants. This demonstrates that the *k*-ABE produced good prediction for such datasets.

Another important raised issue is the impact of feature subset selection (FFS) algorithm on the structure of data, and thereby on the obtained *k*-values. Many research studies in software effort estimation reported the great effect of feature selection on the prediction accuracy of ABE [2, 24]. This paper also investigates whether the use of FSS algorithm can support the proposed method to deliver better predication accuracy. In this paper we used brute-force algorithm that is implemented in ANGEL tool to identify the best features for each dataset. It is recommended to re-apply the adjustment techniques using only the best selected features. This requires applying feature subset selection algorithm [24] prior to building variant methods of *k*-ABE. Although, typically, FSS should be repeated for each training set, it is computationally prohibitive given the large numbers of prediction systems to be built. Instead we performed one FSS for each treatment of each dataset, based on a leave one cross validation and using *MMRE*. This means that the same feature subset is used for all training sets within a treatment and for some of these training sets it will be sub-optimal. However, since a previous study [13] has shown the optimal feature subset varies little with variations in the randomly sampled cases present in the training set, this should have little impact on the results.

The performance figures in Tables 10 and 11 show that although the number of *MMRE* and *pred(0.25)* that has been improved when applying FSS is large, the *MMRE* and *pred(0.25)* differences for each method over a particular dataset are still poor. Generally, The percentage of improvements for *k*-ABE in both *MMRE* and *pred(0.25)* is 83.3%, whilst for *k*-ABE$_{SM}$ is 88.8%, for *k*-ABE$_{GA}$ is 61.1%, and for *k*-ABE$_{NN}$ is 77.7%. However, the significance tests between the *k*-ABE variants with all features and when using only the best features do not show significant differences, so we can conclude that the proposed method works well with all features without the need to apply features subset selection algorithms, and this will reduce computation cost of the whole prediction model especially for large datasets.

Table 10. *MMRE* with Feature Selection results for *k*-ABE variants

| Dataset | *k*-ABE | *k*-ABE$_{SM}$ | *k*-ABE$_{GA}$ | *k*-ABE$_{NN}$ |
|---|---|---|---|---|
| Albrecht | **27.5** | **50.0** | 64.3 | 57.9 |
| Kemerer | **31.6** | **44.5** | 50.8 | **67.5** |
| Desharnais | **31.8** | 42.6 | 47.3 | **64.2** |
| COCOMO | 44.0 | **70.1** | 94.5 | **97.3** |
| Maxwell | **23.7** | **45.3** | 55.2 | 79.5 |
| China | **28.3** | **60.0** | 60.2 | 76.5 |
| Telecom | **30.1** | **57.1** | 40.3 | 73.5 |
| ISBSG | **32.7** | **44.7** | 46.7 | 66.0 |
| NASA | **24.6** | **76.3** | 37.9 | 48.9 |

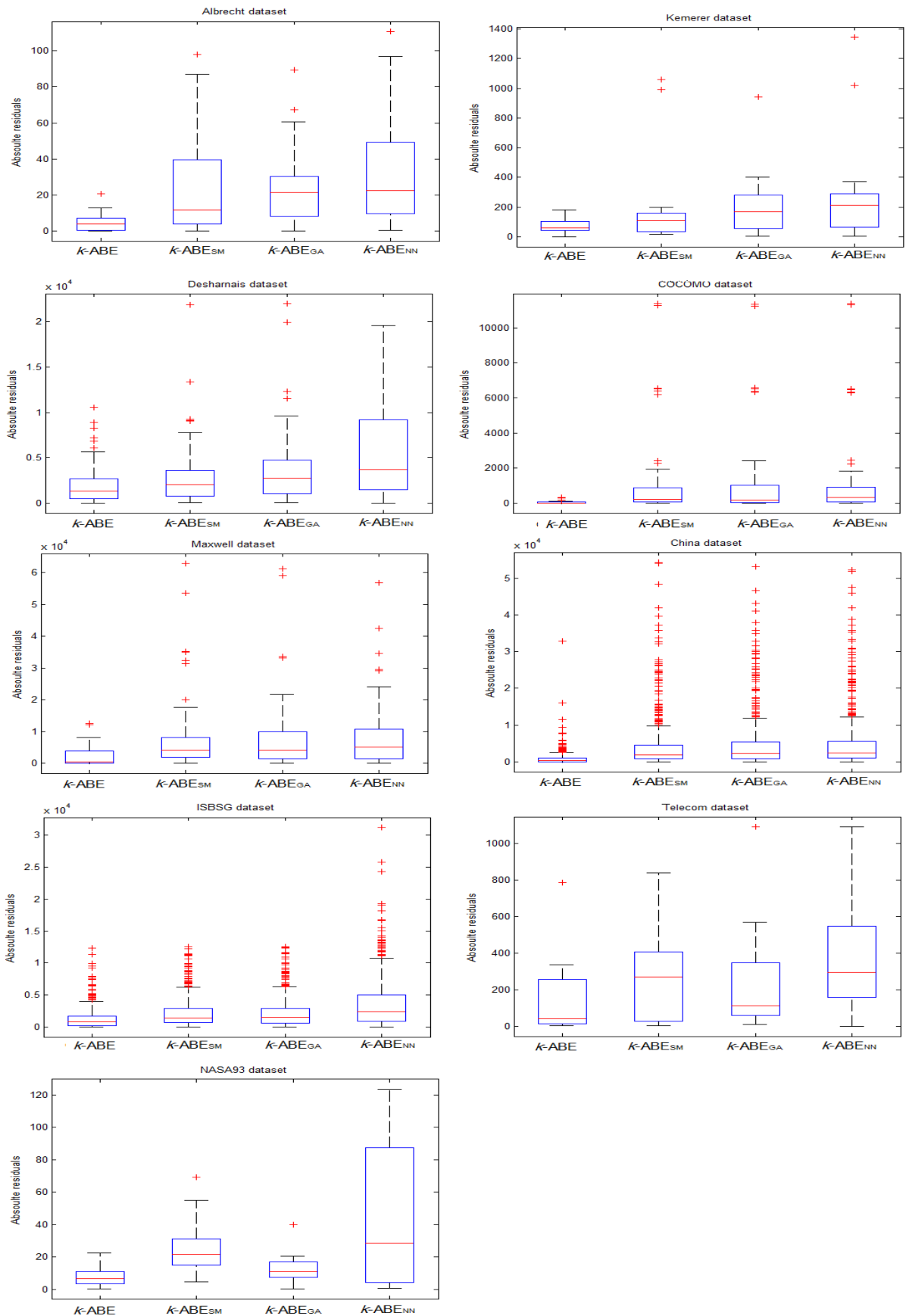

Figure 7. Boxplots of absolute residuals for *k*-ABE variants

Table 11. *pred*(0.25) results with Feature Selection results for *k*-ABE variants

| Dataset | *k*-ABE | *k*-ABE$_{SM}$ | *k*-ABE$_{GA}$ | *k*-ABE$_{NN}$ |
|---|---|---|---|---|
| Albrecht | **48.7** | 20.8 | **29.2** | **25.0** |
| Kemerer | **36.7** | 30.0 | **26.7** | 13.3 |
| Desharnais | **49.0** | 35.6 | 27.3 | 15.6 |
| COCOMO | 38.1 | **18.0** | 9.5 | **10.8** |
| Maxwell | **57.1** | 25.4 | 14.5 | **26.1** |
| China | **48.5** | 14.4 | **16.4** | 10.0 |
| Telecom | 55.6 | 32.2 | 33.3 | **11.1** |
| ISBSG | **41.8** | 27.1 | **32.0** | **20.6** |
| NASA | **46.4** | **30.6** | 27.8 | **33.3** |

To see the predictive performance of *k*-ABE against the most widely used estimation methods in the literature, we compare *k*-ABE with three common methods: Stepwise Regression (SR), Ordinary Least Square Regression (OLS) and Categorical Regression Tree (CART) using the same validation procedure (i.e. leave one cross validation). We have chosen such estimation methods since they use different strategies to make estimate. The remarkable difference between SR and OLS is that OLS generates regression model from all training features whilst SR generates regression model from only significant features. Since some features are skewed and not normally distributed, it is recommended, for SR and OLS, to transform these features using Log transformation such that they resemble more closely a normal distribution [30]. Also, all categorical attributes should be converted into appropriate dummy variables as recommended by [30]. However, all required tests such as normality tests are performed once before running empirical validation which resulted in a general regression model. Then, in each validation iteration a different regression model that resembles general regression model in the structure is built based on the training data set and then the prediction of test project is made on training data set. Table 12 presents a sample of general SR regression models.

Table 12 general Regression models

| Dataset | SR model | $R^2$ |
|---|---|---|
| Albrecht | $Effort = -16.203 + 0.06 \times RawFP$ | 0.90 |
| Kemerer | $Ln(Effort) = -1.057 + 0.9 \times Ln(AdjFP)$ | 0.67 |
| Desharnais | $Ln(Effort) = 4.4 + 0.97 \times Ln(AdjFP) - 1.34 \times L1 - 1.37 \times L2$ | 0.77 |
| COCOMO | $Ln(Effort) = 2.93 - 3.813 \times PCAP + 5.94 \times TURN$ | 0.18 |
| Maxwell | $Ln(Effort) = 633.1234 + 11.273 \times Size$ | 0.71 |
| China | $Ln(Effort) = -2.591 + 4.299 \times Ln(AFP) + 3.151 \times Ln(PDR\_AFP)$ | 0.48 |
| ISBSG | $Ln(Effort) = 5.9318 + 0.261 \times Ln(AFP) + 0.066 \times Ln(ADD)$ | 0.21 |
| Telecom | $Effort = 62.594 + 1.6061 \times changes$ | 0.53 |
| Nasa | $Ln(Effort) = -50.1815 + 34.142 \times Ln(KLOC) - 0.8693 \times ME$ | 0.90 |

As can be seen from Table 12 that the SR model for Desharnais dataset uses the dummy variables *L1* and *L2* instead of the categorical variable (*Dev.mode*). The $R^2$ for COCOMO and ISBSG shows that their SR models were very poor with only 18-21% of the variation in effort being explained by variation in the significant selected features. However, this is not an indicative to the worst of their predictive performance. On the other hand. The log-transformation is used in OLS and SR models to ensure that the residuals of regression models become more homoscedastic, and follow more closely a normal distribution [20]. Tables 13 and 14 show the results from the comparison between *k*-ABE and other regression models: SR, OLS, CART over all datasets. The overall results indicate that the *k*-ABE produces better performance than regression models, but with exception to China and NASA datasets that failed to be superior in terms of *MMRE* and *pred(0.25)*.

Table 13. *MMRE* results of *k*-ABE variants against Regression models

| Dataset | *k*-ABE | SR | CART | OLS |
|---|---|---|---|---|
| Albrecht | **30.5** | 85.6 | 114.45 | 57.3 |
| Kemerer | **38.2** | 38.5 | 96.5 | 63.3 |
| Desharnais | **34.3** | 41.0 | 48.73 | 42.6 |
| COCOMO | **29.3** | 58.3 | 138.74 | 48.8 |
| Maxwell | **27.7** | 73.5 | 57.1 | 75.3 |
| China | 31.6 | **24.2** | 29.82 | 35.5 |
| Telecom | **30.4** | 81.2 | 77.82 | 75.4 |
| ISBSG | **34.2** | 58.0 | 82.9 | 66.4 |
| NASA | 25.6 | **17.1** | 27.6 | 30.5 |

Table 14. *pred*(0.25) results of *k*-ABE variants against Regression models

| Dataset | *k*-ABE | SR | CART | OLS |
|---|---|---|---|---|
| Albrecht | **41.7** | 33.3 | 12.5 | 37.5 |
| Kemerer | 26.7 | **66.7** | 6.7 | 13.3 |
| Desharnais | 40.3 | 39.0 | 36.4 | **43.2** |
| COCOMO | **57.1** | 36.5 | 11.1 | 49.2 |
| Maxwell | 56.5 | 54.2 | **79.2** | 25.8 |
| China | 46.7 | **70.2** | 65.73 | 27.5 |
| Telecom | **55.6** | 38.9 | 38.9 | 27.8 |
| ISBSG | **38.8** | 26.5 | 29.5 | 21.2 |
| NASA | 44.4 | **83.3** | 50.0 | 45.2 |

Table 15 demonstrates the sum of *win*, *tie* and *loss* values that are resulted from the comparisons between *k*-ABE, SR, CART and OLS. Every method is compared to 3 other models, over 3 error measures and 9 datasets, so the maximum value that either one of the *win*, *tie*, *loss* statistics can attain is: *3×3×9 =81*. Notice that the *tie* values are in *12-25* range. Therefore they would not be so informative as to differentiate the methods, so we consult to *win* and *loss* statistics. There is considerable difference between the best and the worst methods in terms on *win* and *loss*. The results show that the *k*-ABE is top ranked method with *win-loss=37* followed by SR in the second place with *win-loss=8*. Interestingly, OLS has the minimum number of *loss* over all datasets.

Table 15. *win-tie-loss* Results for *k*-ABE and other regression models

| ABE variant | *win* | *tie* | *loss* | *win-loss* |
|---|---|---|---|---|
| *k*-ABE | **48** | **22** | **11** | **37** |
| SR | 32 | 25 | 24 | 8 |
| CART | 19 | 23 | 39 | -20 |
| OLS | 5 | 12 | 64 | -59 |

## 6   DISCUSSION AND FINDINGS

This work is an extension of our previous work presented in [18]. In the previous work, the authors tried to learn the *k* analogy value using the Bisecting k-medoid clustering algorithm on historical datasets and they found that there is no static *k* value for all datasets. In spite of this discovery, the previous work has several limitations. First, no adjustment techniques or feature selection methods were used. Furthermore, the effort was estimated using un-weighted mean trained effort of the train projects of the leaf cluster. The limitations of the previous work has been addressed in this extended paper by extending the previous approach so that the optimum *k* value is discovered rather than guessed (as in the previous paper). Moreover, in this new paper, we carried out research to show that the discovered set of analogies work well with different kinds of adjustments techniques such as Similarity Based Adjustment, Genetic Algorithm Based Adjustment and Neural Network Based Adjustment. The main findings of this paper is presented below.

## 6.1 FINDINGS

Based on the obtained results and figures we can summarize our findings as follow:

*Finding 1*: Having seen the bar chart in Figures 1 and 5, there is sufficient believe that the $k$ number of analogies is not fixed and its selection process should take into account the underling structure of dataset as shown in Figure 5. On the other hand, we conjecture that prior reports on discovering $k$ analogies were restricted to limited fixed values staring from 1 to 5. For example Azzeh et al. [2] found that the $k$=1 was the best performer for large datasets and $k$=2 and 3 for small datasets. In contrast, Kirsopp et al. [13] found $k$=2 produced superior results for all employed datasets. Therefore the past believe about finding $k$ value was extremely subject to the human intuition.

*Finding 2*, using many datasets from different domains and sources show that the proposed method has capability to discover their underlying data distribution and automatically come up with optimum set of $k$ analogies for each individual project. The BK method works well with small and large datasets and those that have a lot of discontinuities such as in Maxwell and COCOMO.

*Finding 3*: observing the impact of FSS on the BK method, we can see that our results with FSS are not significantly different than the results without FSS. So they are sufficiently stable to draw a conclusion that FSS is not necessary for any variant of $k$-ABE as they are highly predictive without it.

*Finding 4*: while regular ABE models are deprecated by this study, the simple ABE1 with only one analogy is found to be good choice especially for large datasets. Hence, proponents of this method might elect to explore more intricate form than just simple ABE1.

*Finding 5*: The top ranked method is $k$-ABE confirmed by collecting *win-tie-loss* for each variant of ABE method as shown in Table 16. When we look at the win-tie-loss values in Table 4, we see that in all nine datasets, $k$-ABE has the highest *win−loss* values. This suggests that $k$-ABE has obtained lower *MRE* values than all other methods. Indeed, $k$-ABE has a loss value of 0 for eight datasets (except Maxwell dataset) and this shows that $k$-ABE has never been outperformed by any other method in all datasets for statistically significant cases.

*Finding 6*: Observing *win-tie-loss* results, we can draw a conclusion that the adjustment techniques used in this study do not significantly improve the performance of $k$-ABE variants, hence, the basic $k$-ABE without adjustment is still the most performer model among all variants. This may reduce the computation power needed to perform such estimation especially when number of features and projects is extremely large.

TABLE 16. *win-tie-loss* Values for variants of $k$-ABE and multiple $k$ values over all datasets

| Method | FSS | *win* | *tie* | *loss* | *win-loss* |
|---|---|---|---|---|---|
| $k$-ABE | No | 169 | 40 | 7 | 162 |
| | Yes | 174 | 30 | 12 | 162 |
| $k$-ABE$_{SM}$ | No | 51 | 86 | 79 | -28 |
| | Yes | 46 | 84 | 86 | -40 |
| $k$-ABE$_{GA}$ | No | 43 | 96 | 77 | -34 |
| | Yes | 48 | 75 | 93 | -45 |
| $k$-ABE$_{NN}$ | No | 22 | 76 | 118 | -96 |
| | Yes | 21 | 90 | 105 | -84 |
| ABE1 | No | 68 | 73 | 75 | -7 |
| | Yes | 123 | 66 | 27 | 96 |
| ABE2 | No | 37 | 123 | 56 | -19 |
| | Yes | 104 | 87 | 25 | 79 |
| ABE3 | No | 39 | 114 | 63 | -24 |
| | Yes | 98 | 90 | 28 | 70 |
| ABE4 | No | 41 | 93 | 82 | -41 |
| | Yes | 87 | 87 | 42 | 45 |
| ABE5 | No | 52 | 73 | 91 | -39 |
| | Yes | 92 | 72 | 52 | 40 |

## 6.2 THREATS TO VALIDITY

This section presents the comments on threats to validities of our study based on internal, external and construct validity. Internal validity is the degree to which conclusions can be drawn with regard to configuration setup of BK algorithm including: 1) the identification of initial medoids of BK for each dataset, 2) determining stopping criterion. Currently, there is no efficient method to choose initial medoids so we used random selection procedure. We believe that this decision was reasonable even though it makes the k-medoids is computationally intensive. For stopping criterion we preferred to use the variance performance measure to see when the BK should stop. Although there are plenty of variance measures we believe that the used measure is sufficient to give us indication of how instances in the same clusters are strongly related.

Concerning construct validity which assures that we are measuring what we actually intended to measure. Although there is criticism regarding the used performance measures such as *MMRE* and *pred* [6, 22], we do not consider that choice was a problem because (1) They are practical options for majority of researchers [9, 19, 21], and (2) using such measures enables our study to be benchmarked with previous effort estimation studies.

With regard to external validity, i.e. the ability to generalize the obtained findings of our comparative studies, we used nine datasets from two different sources to ensure the generalizability of the obtained results. The employed datasets contain a wide diversity of projects in terms of their sources, their domains and the time period they were developed in. We also believe that reproducibility of results is an important factor for external validity. Therefore, we have purposely selected publicly available datasets.

## 7 CONCLUSIONS

In this paper, we presented the problem of discovering the optimum set of analogies to be used by ABE in order to make good software effort estimates. However, it is well recognized that the use of fixed number of analogies for all test projects is not sufficient to obtain better predictive performance. In our paper we defined four research questions to address the traditional problem of tuning ABE methods: 1) Understanding the structure of data and 2) Finding a technique to automatically discovering the set of analogies to be used for every single project. Therefore, we proposed a new technique based on utilizing Bisecting k-medoids clustering algorithm and variance degree. Therefore, rather than proposing a fixed *k* value a priori as the traditional ABE methods do, what *k*-ABE does is starting with all the training samples in the dataset, learning the dataset to form BK binary tree and excluding the irrelevant analogies on the basis of variance degree and discovering the optimum set of *k* analogies for each individual project. The proposed technique has the capability to support different size of datasets that have a lot of categorical features. The main aim of utilizing BK tree is to improve the predictive performance of ABE via: 1) building itself by discovering the characteristics of a particular dataset on its own and, 2) excluding outlying projects on the basis of variance degree.

## 8 Acknowledgements

The authors are grateful to the Applied Science University, Amman, Jordan, for the financial support granted to cover the publication fee of this research article.